\documentclass[twocolumn,prm,showpacs,preprintnumbers,amsmath,amssymb,citeautoscript]{revtex4}

\usepackage{graphicx}
\usepackage{dcolumn}
\usepackage{bm}
\usepackage{epstopdf}
\usepackage{textgreek}

\usepackage{color,hyperref}
\definecolor{blue}{rgb}{0.0,0.0,1}
\hypersetup{colorlinks,breaklinks,
            linkcolor=blue,urlcolor=blue,
            anchorcolor=blue,citecolor=blue}

\usepackage{color}

\begin{document}

\preprint{YJ-MC}

\title{Density functional study of native point defects in CaO}

\author{Yunhwa Jo}
\author{Minseok Choi}
\email{minseok.choi@inha.ac.kr}
\affiliation{Department of Physics, Inha University, Incheon 22212, Korea}
\affiliation{Institute of Quantum Science, Inha University, Incheon 22212, Korea}

\date{\today}

\begin{abstract}

We investigate the structural, electronic, and optical properties of native point defects in CaO using first-principles density-functional calculations. Oxygen vacancies are favored under O-poor conditions, whereas calcium vacancies dominate under O-rich conditions. Calculated migration barriers and binding energies indicate that vacancy complexes are thermodynamically stable and can survive high-temperature annealing. Optical transition energies, evaluated using the Franck–Condon framework, suggest that several experimentally observed absorption and emission peaks can be attributed to negatively charged vacancy complexes as well as isolated oxygen vacancies. 
 
\end{abstract}


\maketitle

\section{Introduction}

\textcolor{red}{}


Calcium oxide (CaO) has attracted attention as a potential gate oxide material for metal--oxide--semiconductor structures owing to its wide band gap ($\sim$7.1 eV)~\cite{whited1973exciton} and relatively high dielectric constant ($\sim$11.8)~\cite{albuquerque2008structural}. However, native defects such as oxygen vacancies ($V_{\rm O}$) are known to play a critical role in determining the electrical reliability of CaO-based oxides~\cite{wu2020first,albuquerque2008structural,franckel2025absorption}. Conversely, defects in CaO have also been proposed as promising quantum defects, supported by the long spin coherence time ($T_2$) in CaO, with a reported $T_2$ of 34 ms that exceeds those of many solid-state hosts and is even longer than the typical millisecond-scale values reported for diamond and silicon carbide~\cite{kanai2022generalized}. Davidsson {\it et al.} computationally proposed ${\rm X_{Ca}}$--$V_{\rm O}$ complexes, where X (= Bi, Sb, or I) substitutes on a Ca site and couples with an $V_{\rm O}$, as promising quantum defects because they introduce a two-level electronic structure analogous to that of the NV$^-$ center in diamond~\cite{davidsson2024discovery}.

A few density functional theory (DFT) studies of defects in CaO have been reported. Using the generalized gradient approximation (GGA) and hybrid density functionals, Wu {\it et al.}~\cite{wu2020first} reported that $V_{\mathrm{O}}$ introduces the defect states near mid-gap and suggested that the $V_{\mathrm{O}}$-associated state is responsible for the experimentally observed absorption bands at 340 and 400 nm and emission bands at 370 and 605 nm ~\cite{chen1990effect}. Using time-dependent hybrid functional theory, Franckel {\it et al.}~\cite{franckel2025absorption} suggested that the experimentally observed transitions may originate from neutral ($V_{\mathrm{O}}^{0}$) and singly positive $V_{\mathrm{O}}$ ($V_{\mathrm{O}}^{+}$). Yuan {\it et al.}~\cite{yuan2024first} performed Heyd-Scuseria-Ernzerhof (HSE) hybrid-functional calculations~\cite{heyd2003hybrid,krukau2006influence} and showed that $V_{\mathrm{O}}$ is energetically stable in the 2+ ($V_{\mathrm{O}}^{2+}$) and + ($V_{\mathrm{O}}^{+}$) charge states when the Fermi level lies near mid-gap, whereas $V_{\mathrm{O}}^{0}$ is expected to become stable only when the Fermi level is close to the conduction-band minimum (CBM).


Calcium vacancies ($V_{\rm Ca}$) have also been theoretically investigated as a possible origin of the unusual magnetic properties of CaO~\cite{elfimov2002possible,osorio2006magnetism}. Elfimov {\it et al.} showed that neutral $V_\mathrm{Ca}$ possesses a local magnetic moment of 2~$\mu_B$, which could give rise to half-metallic ferromagnetism. Osorio-Guill{\'e}n {\it et al.}~\cite{osorio2006magnetism} also reported that $V_\mathrm{Ca}$ introduces local magnetic moments associated with shallow acceptor states near the valence-band maximum (VBM). However, they concluded that ferromagnetic coupling between Ca vacancies are short ranged, requiring defect concentrations far in excess of thermodynamic equilibrium to achieve magnetic percolation. More recently, Yuan {\it et al.}~\cite{yuan2024first} reported a HSE study to investigate the intrinsic point defects and how they limit Fermi-level positions and doping in CaO. They showed that while $V_\mathrm{Ca}$ is most stable in the 2$-$ ($V_{\rm Ca}^{2-}$) charge state over most of the band gap, it can also become stable in a positively charged state when the Fermi level is close to the VBM.



Despite these studies, several key aspects of native defects in CaO remain unclear,
including defect complexes, migration barriers, and optical properties. The relative stability between isolated defects and their complexes need to be addressed to provide insight into the dominant defect species in the material. 
Migration barriers govern defect diffusion and are therefore directly related to defect (complex) formation, dissociation, and annealing processes. 



In this work, we systematically performed first-principles DFT calculations to investigate native point defects in CaO. The formation energies of isolated defects and the binding energy of defect complexes were examined to determine prevalent defects. The migration energy of the isolated defects were also evaluated to clarify the defect stability. Then, the optical transition energies associated with defects were examined.

\section{Computational details}
\label{sec:method}

\subsection{Density functional theory}
\label{sec:dft}

DFT calculations were carried out using the projector augmented wave method \cite{blochl1994projector} as implemented in the  {\it Vienna Ab initio Simulation Package} ({\sc vasp}) code \cite{kresse1993ab}. The exchange-correlation energy was described using the Strongly Constrained and Appropriately Normed (SCAN) meta-GGA functional~\cite{sun2015strongly}. The cutoff energy for the plane wave basis was set to 520 eV. The pseudopotentials were used in which 3\emph{s}$^2$3\emph{p}$^6$4\emph{s}$^2$ for Ca, and 2\emph{s}$^2$2\emph{p}$^4$ for O were treated as valence electrons.

For bulk calculations, we used the primitive unit cell of CaO which contains two atoms. The Brillouin zone integration, a 15 $\times$ 15 $\times$ 15 $\Gamma$-centered $k$-point grid and Gaussian smearing method were used and atomic positions were relaxed until the Hellmann-Feynman forces were less than 0.005 eV/\AA. 

Defect calculations were performed using 216-atom supercells constructed from a $3\times3\times3$ expansion of the conventional cell. For the Brillouin zone integrations, a 2 $\times$ 2 $\times$ 2 Monkhorst-Pack scheme was used. The atom coordinates were fully optimized until the Hellmann-Feynman force was less than 0.02 eV/\AA. The effects of spin polarization were also included.

The hybrid functional calculations using the HSE functional \cite{heyd2003hybrid,krukau2006influence} were performed to get correct band-gap to address the optical transition energies associated with defects. The mixing parameter $\alpha$ was set to 0.504 which reproduces the experimental band-gap of 7.09 eV~\cite{whited1973exciton}. Since the calculations demand huge computational cost, one-shot electronic structure calculations were performed using the optimized structures obtained with the SCAN functional. A plane-wave cutoff energy of 520 eV and 216-atom supercells were used, consistent with the SCAN calculations, but to further reduce the computational cost, Brillouin-zone integrations were performed using the $\Gamma$ point only.    
For selected cases, full atomic relaxations were performed to compare our results with those reported in a recent HSE study~\cite{yuan2024first} (Sec.~\ref{sec:comparison}). Atomic positions were relaxed until the Hellmann-Feynman forces were below 0.05 eV/\AA. The 512-atom supercells and the $\Gamma$ point only sampling were used, following the setup in the HSE study~\cite{yuan2024first}.

\subsection{Formation energies and transition levels}

The formation energy of a defect or impurity \emph{X} in charge state of \emph{q} is given by \cite{freysoldt2014first} :
 
 \begin{eqnarray} \label{formE_fomu}
  \begin{split}
   \Delta E_f({X^q}) = E_{\rm tot}({X^q}) - E_{\rm bulk} (\mathrm{CaO}) \\ - \sum\limits_{i}n_i(\mu_i^0 + \mu_i) + {q}{\epsilon_{F}} + \Delta^q 
  \end{split}
   \end{eqnarray}   
where $E_{\rm tot}(\emph{X$^q$})$ is total energy of a supercell containing the defect or dopant \emph{X} in charge state \emph{q}, and $E_{\rm bulk}$(\rm {CaO}) is the total energy of the pristine CaO supercells. ${n_i}$ is the number of atoms of type \emph{i} added (${n_i} > 0$) to or removed (${n_i} < 0$) from the pristine CaO supercells. $\mu$$_i$ are chemical potentials that reflect the abundance of elements in a growth or processing environment, as discussed below. $\epsilon_{F}$ is the Fermi energy, which will be referenced to the VBM. $\Delta$$^q$ is a term that corrects for interactions between supercells and with compensating background charges \cite{kumagai2023finite}.

The charge-state transition level $({q}/{q\prime})$ is defined as the Fermi-level position below which the defect is most stable in charge state ${q}$ and above which the same defect is stable in charge state ${q\prime}$ :

\begin{eqnarray}\label{CTL}
(q/q') &=& \frac {E^{f}[X^q; \epsilon_{F}=0]-E^{f}[X^{q'}; \epsilon_{F}=0]}{q'-q}
\end{eqnarray}
where ${E^f[X^q; \epsilon_{F}} = 0]$ is the formation energy for $X^q$ when $\epsilon_{F}$ is at the VBM. 

\subsection{Atomic chemical potential}

The chemical potentials are variables that reflect the relative abundance of the constituents during growth or processing. The atomic chemical potential $\mu_i$ is referenced to the total energy per atom of the standard phase of the species $\mu_i$: $\mu_\mathrm{Ca}^0$ is the total energy per atom of Ca metal, and $\mu_\mathrm{O}^0$ is half the total energy of an isolated O$_2$ molecule.

The chemical potentials $\mu_i$ ($i$ = Ca and O) are restricted by the formation of limiting phases. They must satisfy $\mu_{\rm Ca} \le 0$, $\mu_{\rm O} \le 0$, and the stability condition of CaO:

\begin{eqnarray} \label{chempo}
 \mu_\mathrm{Ca} + \mu_\mathrm{O} = \Delta{H_f}(\mathrm{CaO})
\end{eqnarray}
where $\Delta{H_f}($CaO$)$ is the formation enthalpy of the CaO crystal. 
Note that the charge-state transition levels are not affected by the choice of chemical potentials.

\subsection{Migration barriers}

Migration barriers were calculated using the climbing-image nudged elastic band (cNEB) method \cite{henkelman2000climbing}. The migration barrier allows us to estimate the temperature at which the impurity becomes mobile. Within transition state theory \cite{vineyard1957frequency}, the rate $\Gamma$ at which an impurity/defect hops to a neighboring equivalent site can be expressed as

\begin{eqnarray} \label{vineyard1957frequency}
 \Gamma = \Gamma_{0}\exp\left(\frac{-E_b}{k_BT}\right)
 \end{eqnarray}
where $k_B$ is the Boltzmann constant and $E_b$ is the calculated migration barrier. The prefactor $\Gamma_0$, related to a typical vibrational frequency, can be approximated as $10^{14}$ s$^{-1}$. An estimate for the annealing temperature $T_a$ at which impurity becomes mobile can then be obtained as the temperature at which the rate $\Gamma$= 1 s$^{-1}$, which leads to $T_a \approx E_b \times 360 \ \mathrm{K/eV}$.

\section{Results and Discussion}
\label{sec:results}

\begin{figure*}[]
\begin{center}
\includegraphics[width = 15cm]{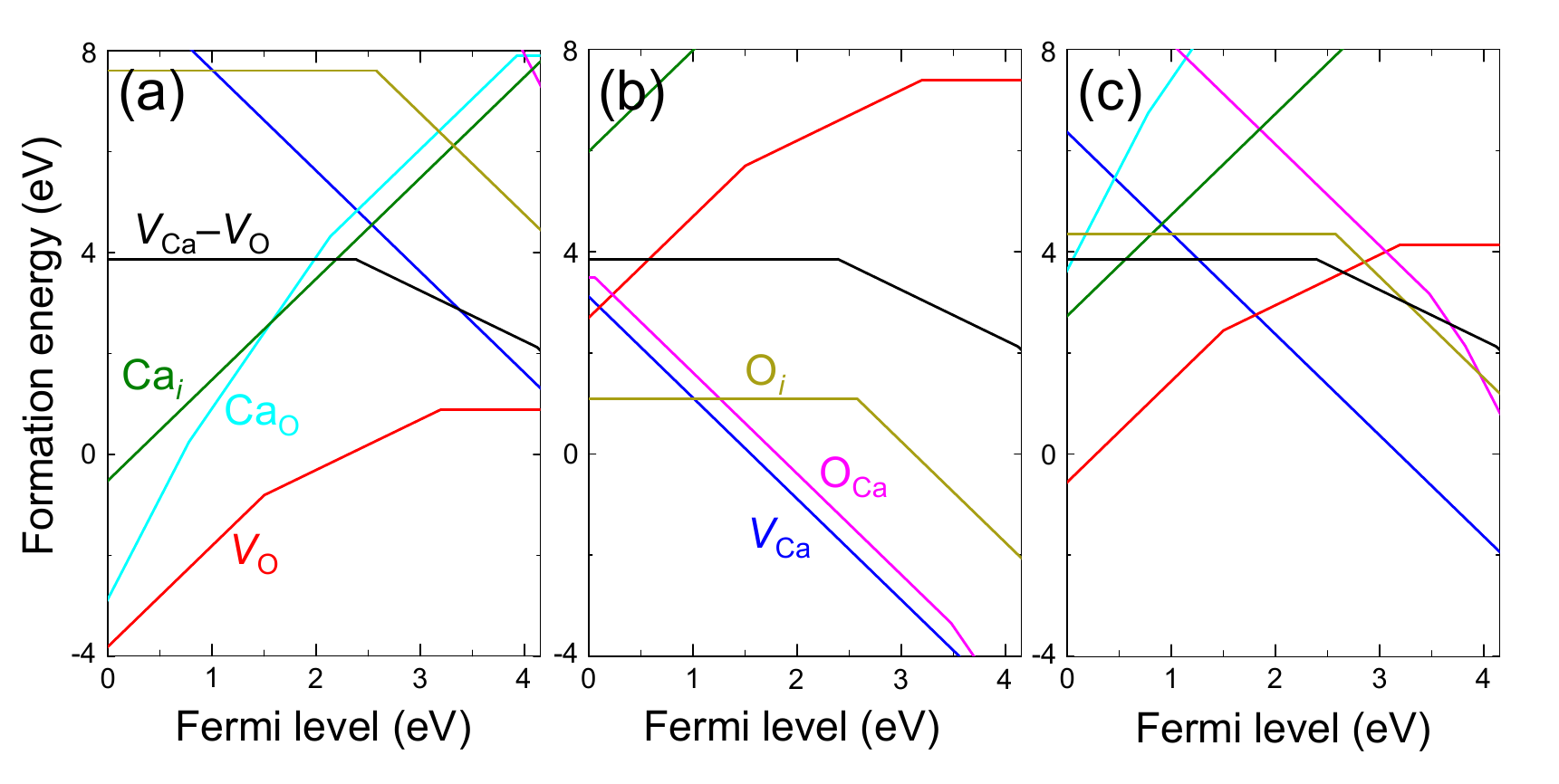}\\
\caption{\label{FormE} Formation energies of native point defects as a function of the Fermi level at (a) O-poor limit, (b) Ca-poor limit, and the condition with the averaged value of $\Delta \mu_ {\rm O}$ between (a) and (b).}
\end{center}
\end{figure*}

\subsection{Bulk properties}
\label{sec:natdef}

\begin{table}[b]
\caption{\label{tab:tab_const} Lattice parameters (lattice constant ($a$), band-gap energy ($E_{\rm g}$), and the formation enthalpy of CaO obtained using the SCAN functional. Experimental values are also shown for comparison.}
\begin{ruledtabular}
\begin{tabular}{ccc}
Property & This work & Experiment  \\
\hline
$a$ (\AA)  & 4.80  & 4.78 \cite{CaO4.78}, 4.81 \cite{CaO4.81_1,CaO4.81_2} \\
$E_{\rm g}$ (eV) & 4.15 & 7.09 \cite{whited1973exciton} \\
$\Delta{H_f}($CaO$)$ (eV) & --6.50 & --6.58 \cite{expform}  \\
\hline
\end{tabular}
\end{ruledtabular}
\end{table}

We first examined the bulk properties of CaO and the results are listed in Table~\ref{tab:tab_const}.
CaO adopts the rocksalt crystal structure with space group $Fm\bar{3}m$, where calcium and oxygen ions form two interpenetrating face-centered cubic sublattices. Each Ca ion is coordinated by six O ions and vice versa, resulting in a highly symmetric octahedral environment. The equilibrium lattice constant is twice the Ca–O bond length in the rocksalt structure. The computed lattice constant is 4.80 \AA, which is close to the experimental values~\cite{CaO4.78,CaO4.81_1,CaO4.81_2}.

The SCAN band gap is calculated to be 4.15~eV, which remains smaller than the experimental value of 7.09~eV~\cite{whited1973exciton}. Nevertheless, it represents an improvement over the values obtained using standard semilocal DFT, reported to be 3.44 and 3.67~eV~\cite{albuquerque2008structural}. The calculated formation enthalpy of CaO is --6.50~eV, also in good agreement with the experimental value of --6.58~eV~\cite{expform}. This value is significantly improved compared with the GGA result of --6.15~eV~\cite{osorio2006magnetism}. These results show that although SCAN still underestimates the band gap, it provides an improved description of both the electronic structure and energetics of CaO compared with standard DFT. 



\subsection{Native defects}

We considered all the possible point defects, including vacancies ($V_{\mathrm{Ca}}$, $V_{\mathrm{O}}$), interstitials (Ca interstitials (Ca$_i$), O interstitials (O$_i$)), antisite-like defects (Ca-on-O defects (${\mathrm{Ca_O}}$), and O-on-Ca defects (${\mathrm{O_{Ca}}}$)). Figure~\ref{FormE} shows the defect formation energies as a function of the Fermi-level position in the band-gap. The slopes of the formation-energy lines indicate the charge state of the defect, and the kinks in the lines correspond to the position of the charge-state transition levels in the band-gap. Lower (higher) formation energies indicate that defects are more likely (unlikely) to form in CaO. 

Overall, donor-like defects, $V_\mathrm{O}$ are the most likely to form at O-poor condition for all the Fermi level position in the gap and two acceptor-like defects, $V_\mathrm{Ca}$ and $\mathrm{O_{Ca}}$ are the most likely at Ca-poor condition for most of the Fermi level positions. However, typical experimental growth conditions lie in between the two conditions, and hence the vacancies would be prevalent. An interesting finding is that the atomic structures of the antisite-like defects resemble interstitial-vacancy complexes.





\subsubsection{Vacancies}
\label{sec:vac}

\begin{figure}[]
\begin{center}
\includegraphics[width = 8cm]{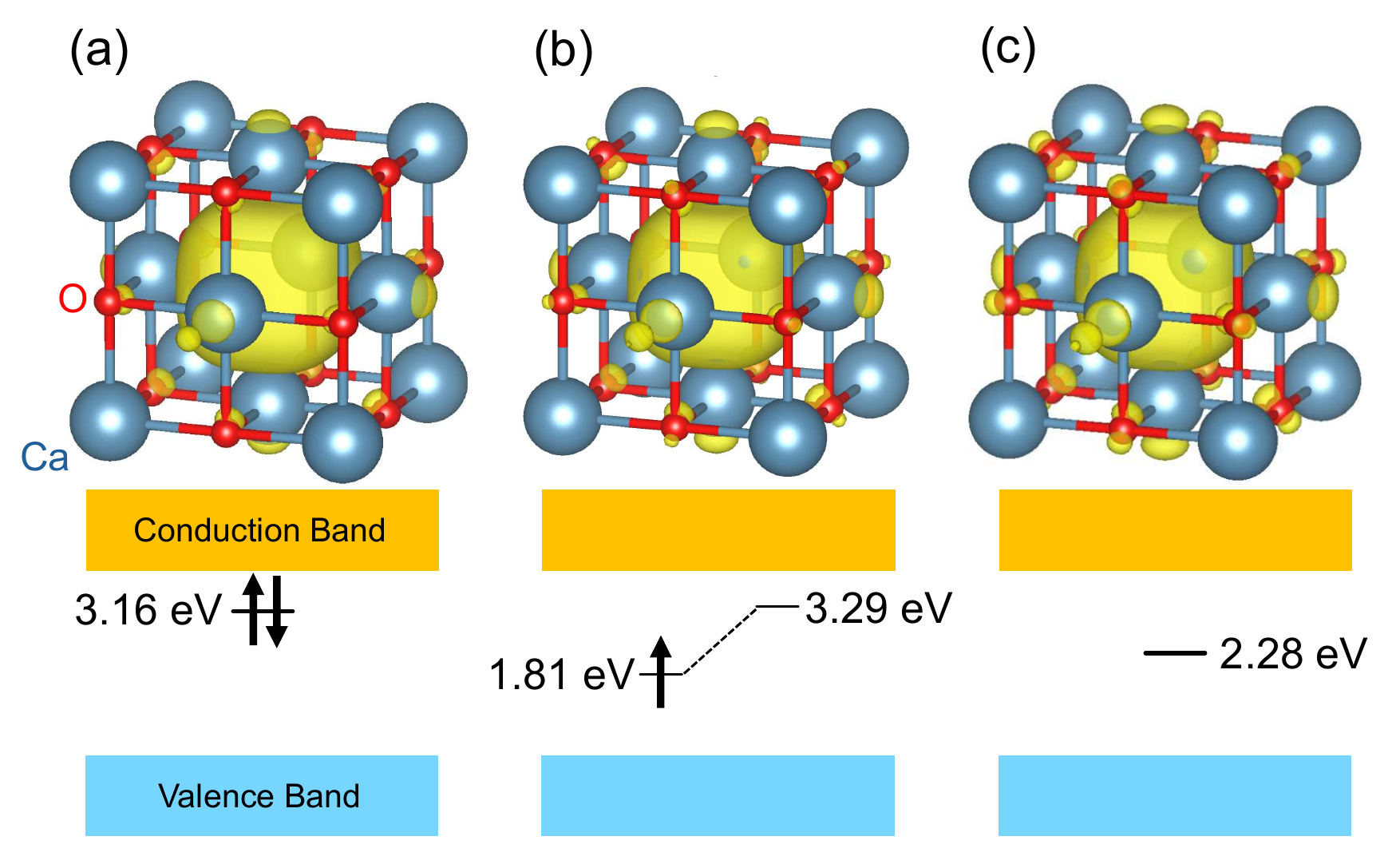}\\
\caption{\label{V_O} Local atomic structure for O vacancies in CaO in (a) the neutral ($V_\mathrm{O}^0$), (b) the + ($V_\mathrm{O}^+$), and (c) the 2+ ($V_\mathrm{O}^{2+}$) charge state. The charge densities of the occupied gap state for $V_\mathrm{O}^0$ and $V_\mathrm{O}^+$ are shown. The isosurfaces correspond to 10\% of the maximum \cite{momma2011vesta}. The corresponding Kohn-Sham states in the band-gap, along with their occupation, are also shown in the lower panels.}
\end{center}
\end{figure}

As shown in Fig ~\ref{FormE}, O vacancies introduce two charge-state transition levels in the band -gap: a (+/0) level at 0.95 eV and (2+/+) level at 2.65 eV below the CBM. $V_\mathrm{O}$ thus acts as a deep donor, which agrees with previous GGA~\cite{wu2020first} and HSE study~\cite{yuan2024first}. 

The charge density of the defect states in the band-gap and the local atomic relaxations are shown in Fig.~\ref{V_O}. Our results show that when $V_\mathrm{O}^0$ forms, the nearest neighboring Ca atoms are displaced outward. In the neutral charge state, the displacement of the Ca atoms is only 0.05 \AA. The Ca atoms relax more significantly in the non-neutral charge states. The Ca atoms relax outward by 0.12 \AA~ in the + charge state and by 0.20 \AA~ in the 2+ charge state.

The in-gap Kohn-Sham states are 
positioned at 3.16 eV for $V_\mathrm{O}^0$, 1.81 eV (electron occupied) and 3.29 eV (electron unoccupied) for $V_\mathrm{O}^+$, and 2.28 eV for $V_\mathrm{O}^{2+}$ above the VBM. $V_\mathrm{O}^+$ has spin 1/2 where an unpaired electron exists. Thus, the defect may contribute to magnetic noise and limit the coherence time of spin qubits in CaO.


Using the cNEB method, the migration barriers were obtained to be $E_b$ = 1.88 eV for $V_\mathrm{O}^{2+}$ and 3.16 eV for $V_\mathrm{O}^{+}$. The corresponding $T_a$ are 677 K and 1138 K, respectively, hence the defects may not be mobile in the samples at room temperature. Previous experimental studies demonstrated that CaO remains structurally stable upon vacuum annealing up to approximately 1000 K, where improved crystallinity was observed \cite{cui2015evolution}. This is consistent with high $T_a$ for O vacancies. 


Regarding $V_\mathrm{Ca}$, our calculations show atypical aspects unlike other studies. The removal of a Ca atom leads to 2-- charge state only \cite{elfimov2002possible,osorio2006magnetism}. I.e., $V_\mathrm{Ca}$ is thermodynamically stable in the 2-- charge state for any Fermi-level position in the band-gap. No in-gap defect state is found. 

\begin{figure}[]
\begin{center}
\includegraphics[width = 6cm]{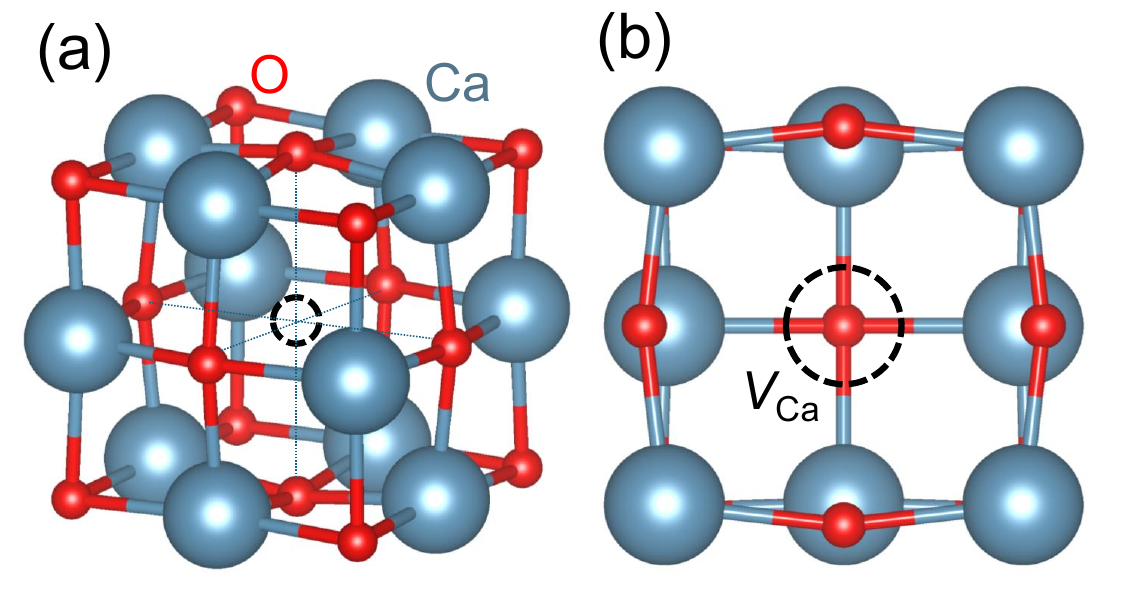}\\
\caption{\label{V_Ca} (a) Local atomic structure and (b) side veiw for Ca vacancies in the 2- ($V_\mathrm{Ca}^{2-}$) charge state.}
\end{center}
\end{figure}

Figure~\ref{V_Ca} shows local atomic structure of $V_\mathrm{Ca}^{2-}$. The neighboring six O atoms move outward by 0.19 \AA. Note that when some of the O atoms were manually moved to break the local symmetry, they move back to the original positions, and we could not find the other charge state and atomic configurations which are energetically stable.

The calculated $E_b$ for $V_\mathrm{Ca}^{2-}$ is 2.09 eV using the cNEB method, which corresponds to $T_a$ = 752 K. Thus, $V_{\rm Ca}^{2-}$ is not very mobile, similar to oxygen vacancies.

We further examined the stability of the $V_\mathrm{Ca}$--$V_\mathrm{O}$ complexes. In practice, the growth conditions are expected to lie between the O-rich (Ca-poor) and O-poor (Ca-rich) limits. To mimic realistic growth conditions, we set $\Delta\mu_{\rm O}$ to the average of the O-rich and O-poor limits. As shown in Fig.~\ref{FormE}(c), both $V_\mathrm{Ca}$ and $V_\mathrm{O}$ have relatively low formation energies and are therefore likely to form. Thus, we examined the stability of vacancy complexes composed of the most stable charge states of the constituent vacancies: $V_\mathrm{Ca}^{2-}$, $V_\mathrm{O}^{+}$, and $V_\mathrm{O}^{2+}$.

\begin{figure}[b]
\begin{center}
\includegraphics[width = 6cm]{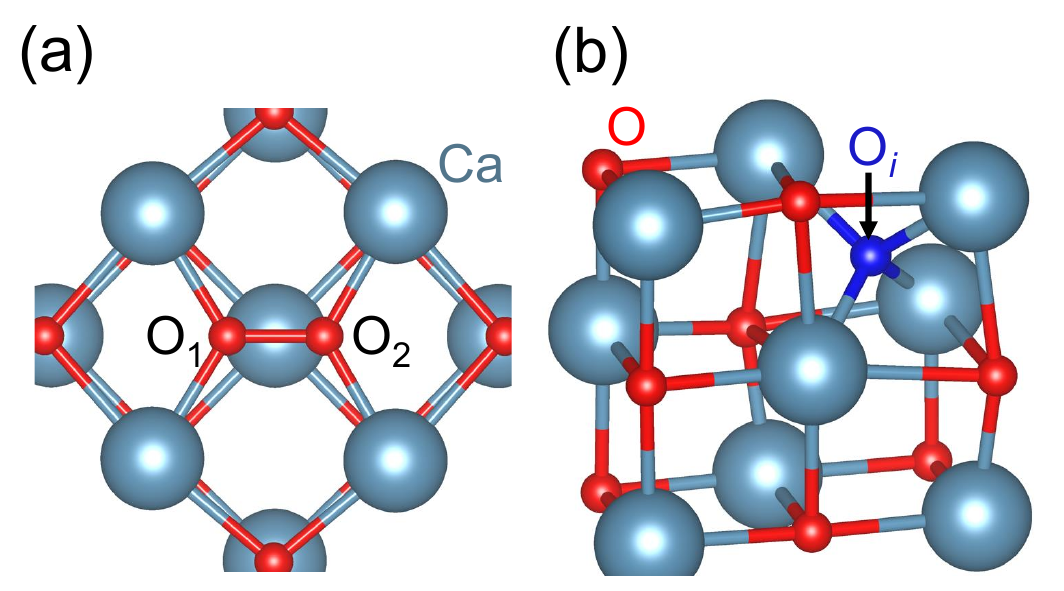}\\
\caption{\label{Oi} Local atomic structure for O interstitials in (a) the neutral (O$_i^0$), (b) the 2-- (O$_i^{2-}$) charge state.} 
\end{center}
\end{figure}

The binding energy of $(V_\mathrm{Ca}-V_\mathrm{O})^{0}$ is calculated as $E_{\rm bind}[(V_\mathrm{Ca}-V_\mathrm{O})^{0}] = E^f(V_\mathrm{Ca}^{2-})+E^f(V_\mathrm{O}^{2+})-E^f[(V_\mathrm{Ca}-V_\mathrm{O})^{0}]$, resulting in a value of 1.95 eV. The positive binding energy indicates that the complex formation is favorable. $E_{\rm bind}[(V_\mathrm{Ca}-V_\mathrm{O})^{-}] = E^f(V_\mathrm{Ca}^{2-})+E^f(V_\mathrm{O}^{+})-E^f[(V_\mathrm{Ca}-V_\mathrm{O})^{-}]$ gives the value of 1.07 eV, indicating that the formation of the complexes is also favorable. The $(V_\mathrm{Ca}-V_\mathrm{O})^{-}$ complex is likewise paramagnetic, suggesting that these defects may act as sources of magnetic noise in CaO-based quantum defects. 

The dissociation energy can be estimated by adding the migration barrier for the mobile defect species to the binding energy of the complexes. Using the lowest migration barrier among  $V_{\rm Ca}^{2-}$ and $V_{\rm O}^{2+}$ ($V_{\rm O}^{+}$), we would estimate an activation energy of $E_{\rm bind}$  + $E_b$ = 1.95 eV + 1.88 eV = 3.83 eV for $(V_\mathrm{Ca}-V_\mathrm{O})^{0}$ and 1.07 eV+ 2.09 eV = 3.16 eV for $(V_\mathrm{Ca}-V_\mathrm{O})^{-}$, which roughly provides an estimate for the annealing temperature of 1379 K and 1138 K, respectively.




\subsubsection{Interstitials}
\label{sec:inter}

The oxygen interstitial, O$_i$, has modest formation energies at O-rich limit (Fig.~\ref{FormE}(b)). However, the defects may form only when the Fermi level is close to the conduction band. The defect has a (0/2--) level at 2.29 eV above the VBM, which indicates that the interstitials are deep acceptors. 

O$_i$ can have two distinct configurations. In the neutral charge state, O$_i$ forms as a split-interstitial, as shown in Fig ~\ref{Oi}(a). The bond length of an $\mathrm{O}_1-\mathrm{O}_2$ bond is 1.46 \AA, which is similar to those in other oxides such as ZnO (1.51~\AA)~\cite{AJ_ZnO_2007} and CaWO$_4$ (1.43~\AA) ~\cite{MC_CWO_2026}. In the 2-- charge state, the $\mathrm{O}_i-\mathrm{O}$ aligned corner to corner with an interatomic distance of 2.51 \AA. The $\mathrm{O}_i$ bonds to four neighboring Ca atoms with a bond length of 2.07 \AA.


The computed migration barrier is 0.13 eV for $\mathrm{O}_i^0$ and 0.56 eV for $\mathrm{O}_i^{2-}$. The corresponding $T_a$ are 50 K and 202 K, respectively, hence the interstitials are highly mobile and unlikely to persist as isolated defects.



\begin{figure}[]
\begin{center}
\includegraphics[width = 6cm]{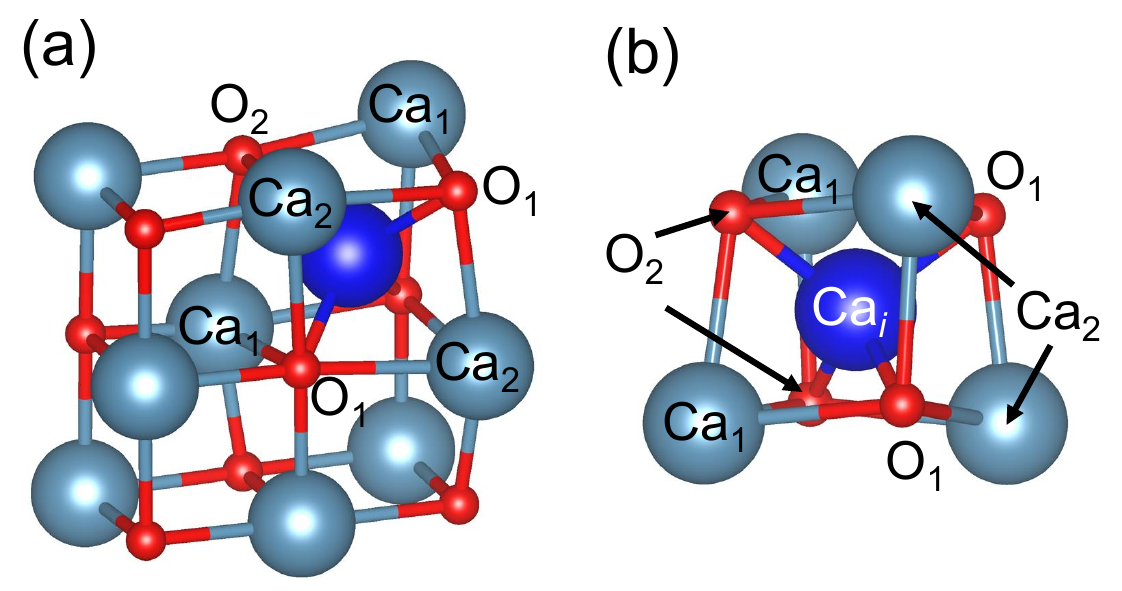}\\
\caption{\label{Cai} (a) Local atomic structure for Ca interstitial in 2+ charge state ($\mathrm{Ca}_i^{2+}$) and (b) another view around Ca$_i$.}
\end{center}
\end{figure}

The calcium interstitial, Ca$_i$, has high formation energy and is stable in the 2+ charge state only, except for the Fermi level position near the valence band, even at Ca-rich limit. Therefore, the defects are unlikely to form in CaO if the samples are obtained at growth conditions close to thermal equilibrium.


Figure~\ref{Cai} shows the local atomic structure of Ca$_i$ that is surrounded by four O and four Ca atoms. Ca$_i$ bonds to four O atoms. The bond lengths are 2.11 \AA~ with two O$_1$ and 2.12 \AA~ with two O$_2$. The interatomic distances to the four neighboring Ca atoms are 2.55 \AA~ for Ca$_1$, two Ca$_2$ and 2.54 \AA~ for Ca$_2$. 

The computed migration barrier is 0.62 eV for $\mathrm{Ca}_i^{2+}$. The corresponding $T_a$ is 223 K. Hence the interstitial is highly mobile and unlikely to be present at isolated defects, like as O interstitials.




\subsubsection{Antisite-like defects}
\label{sec:anti}

\begin{figure}[]
\begin{center}
\includegraphics[width = 8cm]{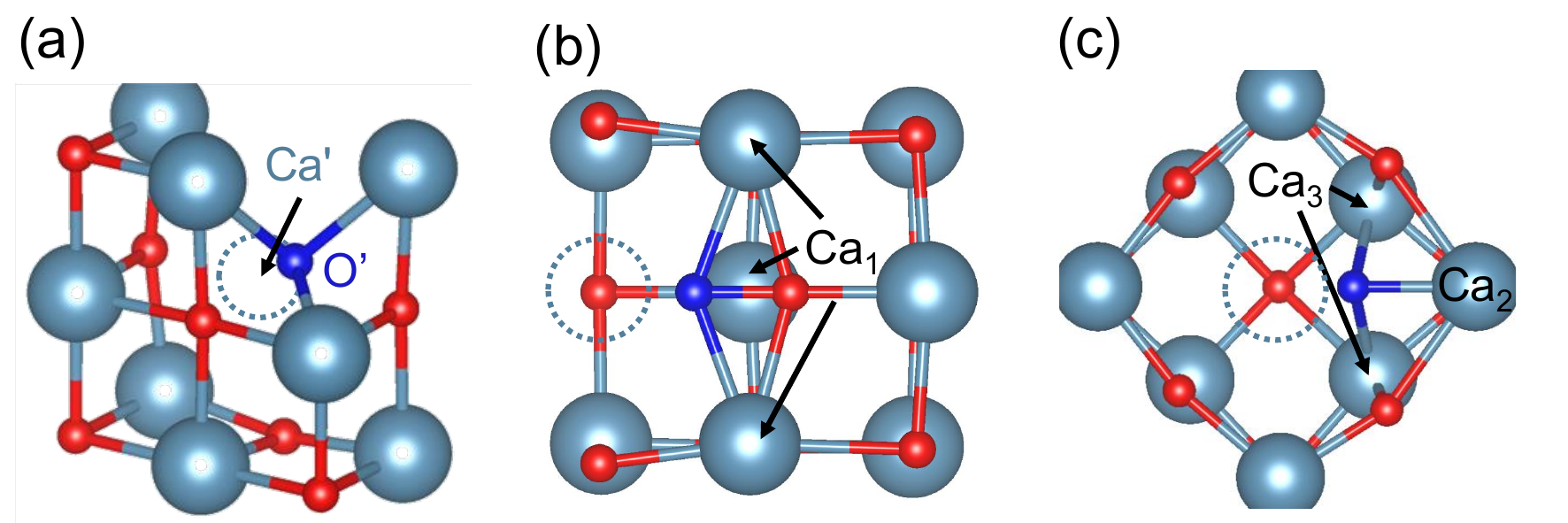}\\
\caption{\label{OCa} Local atomic structures for $\mathrm{O_{Ca}}$ in (a) the neutral (O$_\mathrm{Ca}^{0}$), (b) the 2-- (O$_\mathrm{Ca}^{2-}$), and (c) the 3-- (O$_\mathrm{Ca}^{3-}$) charge state. The structure in the 4 -- charge state is quite similar to that in the 3-- charge state (see text).}
\end{center}
\end{figure}

Although $\mathrm{O_{Ca}}$ has low formation energies under Ca-poor condition (Fig~\ref{FormE} (b)), the defect may be not likely to form in experiment since its formation energies are quite high as seen in Fig~\ref{FormE} (c). The defect has a (0/2--) level at 0.05 eV, (2--/3--) level at 3.42 eV, and (3--/4--) level at 3.71 eV above the VBM.


$\mathrm{O_{Ca}}$ shows significant offcentering (Fig.~\ref{OCa}). $\mathrm{O_{Ca}}$ was examined by putting an O atom (O') at the Ca site (Ca'), but the O' is largely displaced from the Ca'. Indeed, one may consider ${\rm O}_i-V_{\rm Ca}$ complexes instead of $\mathrm{O_{Ca}}$. In the neutral charge state, O' atom moves by  1.23 \AA~ from the Ca' to the corner and forms bonds with the three nearest Ca atoms. The O'--Ca bond length is 2.53 \AA~. In the 2-- charge state, O' moves by 1.42 \AA~ from the Ca' to the face center and makes an O--O like bond. Indeed, the O'--O bond length is 1.48 \AA, which is quite close to that of O$_i$ split-interstitials in Sec.~\ref{sec:inter}. The O'--Ca bond length with neighboring four Ca (Ca$_1$) atoms is 2.48 \AA. Such an O--O like bond is not found in the 3-- and 4-- charge states, instead O' forms a pyramid-like structure with five Ca atoms: one shorter bond with Ca$_2$ and four longer bonds with Ca$_3$. The O'-Ca bond lengths are 2.16 \AA~ for Ca$_2$ and 2.70 \AA~ for Ca$_3$ and the distance between O' and Ca' is 1.24 \AA~ in the 3-- charge state, 2.09 {\AA} for Ca$_2$, 2.43 \AA~ for Ca$_3$ and 1.51 \AA~for Ca' in the 4-- charge state.

\begin{figure}[b]
\begin{center}
\includegraphics[width = 6cm]{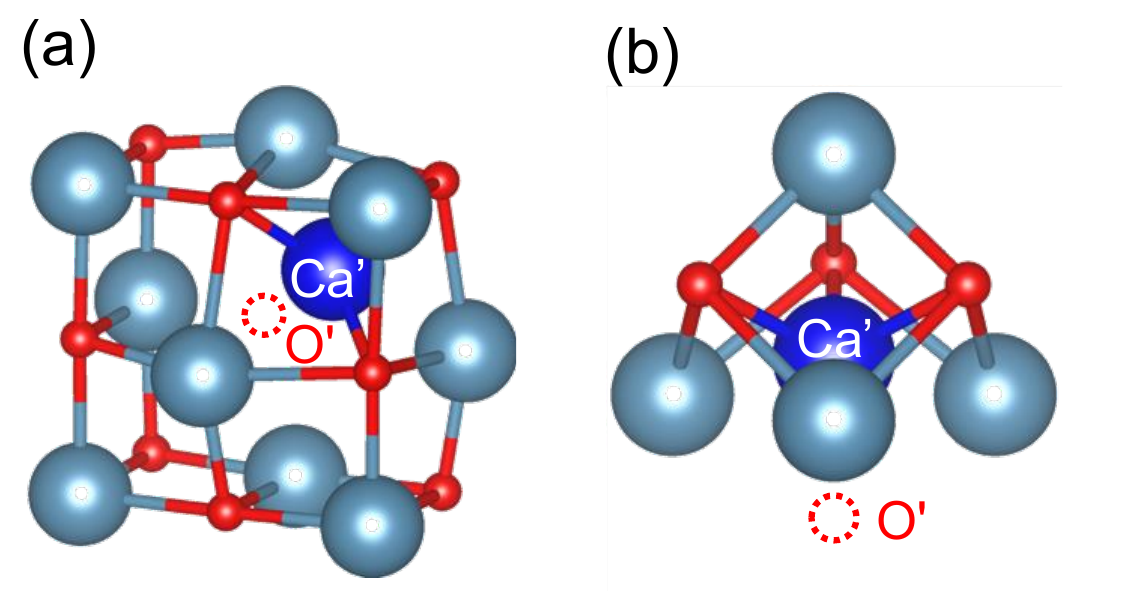}\\
\caption{\label{Ca_O} (a) Local structure for Ca$_\mathrm{O}$ in the neutral charge state ($\mathrm{Ca}_\mathrm{O}^0$) and (b) its another view. The defect has the same local symmetry in the other charge states.}
\end{center}
\end{figure}

\begin{figure*}[]
\begin{center}
\includegraphics[width = 15cm]{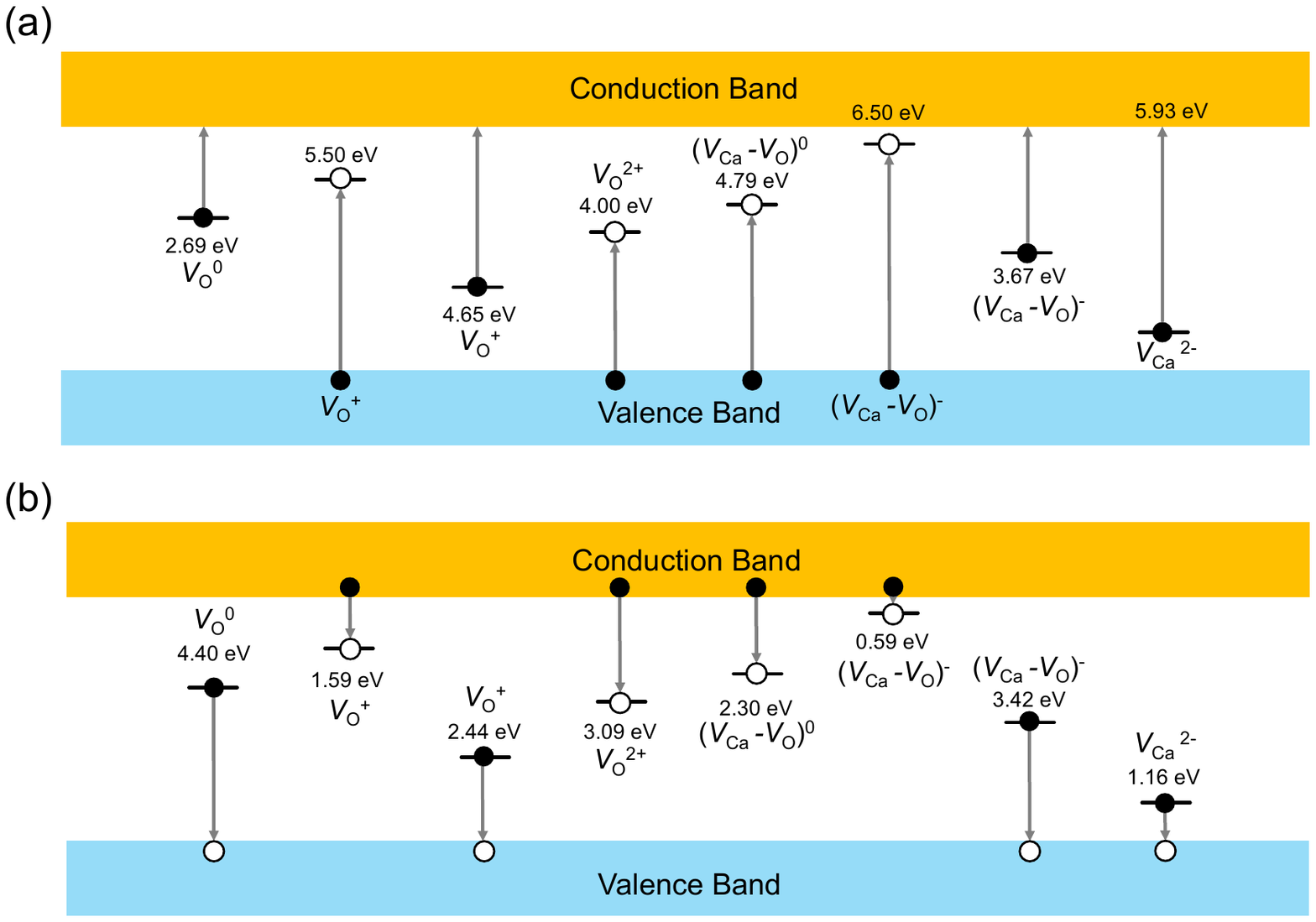}\\
\caption{\label{opti} (a) Optical absorption and (b) emission associated with native defects in CaO. The depicted levels are not zero-phonon lines but correspond to peaks in absorption or emission spectra. Note that $V_{\rm Ca}^{2-}$ introduces 0.05 eV above the VBM when the HSE ($\alpha$ = 0.504) is applied.}
\end{center}
\end{figure*}
We examined the stability of $\mathrm{{O}_{Ca}^{2-}}$ as ${\rm O}_i-V_{\rm Ca}$ complexes by evaluating a binding energy. The corresponding binding energy can be given by $E_{\rm bind}(\mathrm{O_{Ca}^{2-}}) = E^f(V_\mathrm{Ca}^{2-})+E^f(\mathrm{O}_{i}^{0})-E^f(\mathrm{O_{Ca}^{2-}})$. The calculated binding energy is 0.60 eV, indicating that the complex is preferred over separated $V_\mathrm{Ca}$ and O$_i$. But, the estimated dissociation energy of 0.60 eV + 0.13 eV ($E_b$ for $\mathrm{O}_{i}^{0}$) = 0.73 eV, corresponding to $T_a$ is 262 K. This modest binding energy indicates that the complex may not form during growth.  



Ca$_\mathrm{O}$ has the (4+/3+) level at 0.78 eV, (3+/2+) at 2.13 eV, and (2+/0) level at 3.95 eV above the VBM (Fig.~\ref{FormE}). However, the defect has high formation energy even at O-poor limit, indicating that it is not likely to form in the sample grown under thermodynamic equilibrium. 

Similar to O$_\mathrm{Ca}$, ${\rm Ca}_{\rm O}$ is better viewed as a ${\rm Ca}_i$--$V_{\rm O}$ complex due to its significant off-centered structures (Figure~\ref{Ca_O} ). $\mathrm{Ca_{O}}$ was initially examined by putting a Ca atom (Ca') at the O site (O'), but the Ca' is largely displaced from the O'. The Ca' atom is largely displaced from the O' by 1.83 \AA~ in the neutral, 2.02 \AA~ in the 2+, 1.71 \AA~ in the 3+, and 1.40 \AA~ in the 4+ charge state. This displacement allows the Ca atom to bond with three neighboring O atoms and their bond lengths are identical: 2.23 \AA~ in the neutral, 2.11 \AA~ in the 2+, 2.12 \AA~ in the 3+, and 2.14 \AA~ in the 4+ charge state.

The stability of $\mathrm{Ca_{O}}$ was examined as ${\rm Ca}_i-V_{\rm O}$ complexes by evaluating a binding energy. The binding energy  $E_{\rm bind}(\mathrm{Ca_{O}^{2+}}) = E^f(V_\mathrm{O}^{0})+E^f(\mathrm{Ca}_{i}^{2+})-E^f(\mathrm{Ca_{O}^{2+}})$ is 0.33 eV. Thus, the estimated dissociation energy is 0.33 eV + 0.62 eV ($E_b$ for $\mathrm{Ca}_{i}^{2+}$) = 0.95 eV, corresponding to $T_a$ is 342 K. This modest binding energy indicates that the complex may not form during growth. The calculated binding energy $E_{\rm bind}(\mathrm{Ca_{O}^{3+}}) = E^f(V_\mathrm{O}^{+})+E^f(\mathrm{Ca}_{i}^{2+})-E^f(\mathrm{Ca_{O}^{3+}})$ is --0.74 eV, indicating that the complex formation is unfavorable.







\subsubsection{Comparison with hybrid functional}
\label{sec:comparison}


To further examine the reliability of the predicted SCAN results, we compared the SCAN results with hybrid-functional calculations. A recent HSE study employing a tuned mixing parameter reported some unusual aspects of defect charges in CaO. For example, although Ca vacancies are well known acceptors, the HSE predicts thermodynamically stable neutral and positively charged Ca vacancies~\cite{yuan2024first}. We were also able to reproduce the results using the HSE with a tuned mixing parameter ($\alpha$=0.504). In contrast, our SCAN calculations predict that the positive charge states are not thermodynamically stable, which is consistent with the typical feature of Ca vacancies in oxides~\cite{osorio2006magnetism, MC_CWO_2026} (See Sec.~\ref{sec:vac}).

To investigate the origin of this discrepancy, we additionally performed HSE calculations using two different mixing parameters ($\alpha=0.25$ and 0.504). As the fraction of exact exchange decreases, the defect energetics change substantially: the (+/0) charge-transition level shifts from 1.35 eV ($\alpha=0.504$) to 0.32 eV above the VBM ($\alpha=0.25$), while the (0/2--) level decreases from 1.36 to 0.44 eV above the VBM. These results demonstrate that the predicted stability of $V_{\mathrm{Ca}}^{0}$ and $V_{\mathrm{Ca}}^{+}$ is highly sensitive to the choice of the mixing parameter.

This pronounced dependence suggests that reproducing the experimental band-gap alone is not sufficient to guarantee quantitatively reliable defect energetics. Alkauskas \textit{et al.}~\cite{alkauskas2008defect} showed that charge-transition levels of atomically localized defects generally remain nearly unchanged between semi-local and hybrid density functionals once the electronic structures are properly aligned to a common external reference potential, despite substantial changes in the gap. They further demonstrated that an accurate description of defect energetics requires not only the correct band gap but also an appropriate alignment of the band edges with respect to the average electrostatic potential (or equivalently the vacuum level)~\cite{alkauskas2011bandedge}. Therefore, the agreement of the band-gap with experiment does not by itself ensure the accuracy of calculated charge-transition levels.

The stability of the positive charge states may originate from changes in the absolute positions of the band edges, an enhanced localization of the defect state with increasing exact exchange, or a combination of both effects. Although the present calculations do not allow us to distinguish quantitatively between these mechanisms, they clearly indicate that the stabilization of positively charged Ca vacancies predicted by the band-gap-tuned HSE functional should be interpreted with caution. In this respect, the SCAN functional provides a self-consistent description of the defect energetics without empirical tuning of the exchange fraction and offers an alternative picture for the charge-state stability of Ca vacancies in CaO.

\subsection{Optical properties}

Finally, the optical transition peaks possibly associated with defects in experiment are addressed. Experimentally, the optical peaks have been observed in CaO crystals at 340 nm (3.65 eV) and 400 nm (3.11 eV) for absorption and the peaks at 370 nm (3.35 eV) and 605 nm (2.05 eV) for emission~\cite{chen1990effect}. 
To obtain reliable optical transition energies, we used SCAN to determine the stable defect configurations and thermodynamically stable charge states, and subsequently evaluated the optical transition energies using one-shot HSE calculations within the Franck-Condon framework. In this procedure, the SCAN-relaxed geometries and SCAN-predicted stable charge states were retained, while HSE was used only to correct the electronic transition energies. This choice is motivated by the strong dependence of the HSE defect energetics on the mixing parameter, for which the band-gap-tuned HSE functional stabilizes charge states that are not stable in SCAN. 


Because CaO is a wide-band-gap material, strong n-type or p-type doping is expected to be difficult in nominally undoped samples. We therefore focus on defect species that are stable or have low formation energies near the middle of the gap. 
In addition, the experimental growth conditions are expected to lie between the O-rich and O-poor limits. Based on our results, $V_\mathrm{O}^+$, $V_\mathrm{O}^{2+}$, $V_\mathrm{Ca}^{2-}$, and their complexes $(V_\mathrm{Ca}-V_\mathrm{O})^0$ and $(V_\mathrm{Ca}-V_\mathrm{O})^-$ are plausible near the mid-gap (Fig.~\ref{FormE}(c)).
The results are summarized in Fig.~\ref{opti}.


Our results show that the transition $(V_\mathrm{Ca}-V_\mathrm{O})^- \rightarrow (V_\mathrm{Ca}-V_\mathrm{O})^0 + e^-$ gives an absorption energy of 3.67 eV. This agrees well with the experimental absorption peak at 3.65 eV (340 nm) that was assigned to $F^+$-center peak~\cite{chen1990effect}. The energies for $V_\mathrm{O}^+$-mediated absorption are 4.65 eV and 5.50 eV, which are close to the values of 4.73 eV and 5.07 eV in a recent hybrid functional study~\cite{franckel2025absorption}, while these are far from the measured peak at 3.65 eV. These results suggest that the absorption peak may be more naturally attributed to vacancy complexes than to isolated O vacancies. 

Chen {\it et al.}~\cite{chen1990effect} also reported an absorption band at 260~nm (4.77~eV) in thermochemically reduced CaO crystals and attributed it to an unidentified electron-donor defect. Our calculations predict an optical transition at 4.79~eV for $(V_{\mathrm{Ca}}-V_{\mathrm{O}})^{0}$, which is quite close to the experimental value. This agreement allows us to assign the previously unresolved 260~nm absorption band to $(V_{\mathrm{Ca}}-V_{\mathrm{O}})^{0}$.

The transition $V_\mathrm{O}^0 \rightarrow V_\mathrm{O}^+ + e^-$ yields an ionization energy of 2.69 eV. This value lies in the same spectral range as the experimental peak at 3.11 eV (400 nm), previously assigned to the $F$-center absorption~\cite{chen1990effect}, and is also comparable to the hybrid-functional value of 3.22 eV reported recently~\cite{franckel2025absorption}. However, assigning this peak to a thermodynamically stable, isolated $V_\mathrm{O}^0$ requires caution because our calculations show that $V_\mathrm{O}^0$ is stable only when the Fermi level lies close to the CBM.

The temperature dependence of the experimentally observed $F$-center signal provides further insight. The $F$-center-related peak decreases with increasing temperature~\cite{Henderson_1972,Welch_1980}, and Welch {\it et al.} reported that above $T \sim 100$ K, an electron in the excited $F$ center is released into the conduction band through a thermally assisted process with an activation energy of $\sim$ 0.1 eV~\cite{Welch_1980}. This small activation energy can be rationalized by treating the electron in the optically excited state of the $F$ center, which is diffuse and hydrogenic-like, as bound to $V_{\rm O}^+$. The corresponding binding energy is $E_{\rm donor}=13.6~{\rm eV}\times \frac{m^*/m_e}{\varepsilon_{\rm s}^2},$ where $\varepsilon_{\rm s}=11.96$ is the static dielectric constant of CaO~\cite{subramanian1989dielectric} and $m^*=1.22,m_e$ is the electron effective mass at the CBM~\cite{Medvedeva2007}. This estimate gives $E_{\rm donor}\sim 0.12$ eV, which agrees well with the experimentally reported activation energy of $\sim 0.1$ eV~\cite{Welch_1980}.

This quantitative agreement suggests that the $V_\mathrm{O}^0$-related absorption is more appropriately interpreted as arising from a hydrogenic donor-like state, namely $V_\mathrm{O}^{+}$ weakly binding an electron, rather than from a thermodynamically stable, isolated $V_\mathrm{O}^0$ defect. Upon thermal ionization of this weakly bound electron, the population of $V_\mathrm{O}^{+}$ increases, which is consistent with the experimentally observed enhancement of the $F^+$-center-related peaks at elevated temperature~\cite{Henderson_1972}.

For the emission, the transition $(V_\mathrm{Ca}-V_\mathrm{O})^0 + e^- \rightarrow (V_\mathrm{Ca}-V_\mathrm{O})^-$ corresponds to the energy of 2.30 eV. The value is close to the experimental peak at 2.05 eV (605 nm) that was assigned to $F$-center peak~\cite{chen1990effect}. None of $V_\mathrm{O}^0$-related emission process gives the energies close to 2.05 eV. The transition energy for $V_\mathrm{O}^0 + h^+ \rightarrow V_\mathrm{O}^{+}$ is 4.40 eV, and the transition energy for $V_\mathrm{O}^+ + e^- \rightarrow V_\mathrm{O}^{0}$ is 1.59 eV. 

The transition $V_\mathrm{O}^{2+} + e^- \rightarrow V_\mathrm{O}^{+}$ gives 3.09 eV and the transition $(V_\mathrm{Ca}-V_\mathrm{O})^- + h^+ \rightarrow (V_\mathrm{Ca}-V_\mathrm{O})^0$ provides 3.42 eV. These are comparable with another experimental emission peak at 3.35 eV (370 nm), assigned to $F^+$-center peak~\cite{chen1990effect}. 
The transition $V_\mathrm{O}^+ + h^+ \rightarrow V_\mathrm{O}^{2+}$ gives the energy of 2.44 eV, which is far from the experimental value. 





\section{SUMMARY}

We have investigated the structural and electronic properties of native point defects in CaO using density functional theory and the SCAN functional. Oxygen vacancies are favored under O-poor conditions, whereas calcium vacancies dominate under O-rich conditions. In experiment, both the vacancies may be likely to form since the sample growth conditions would be in between O-poor and O-rich conditions. The calculated migration barriers indicate that isolated vacancies are immobile at room temperature but become mobile during high-temperature annealing. Favorable binding energies further show that the formation of oxygen--calcium vacancy complexes is thermodynamically favorable in experiments.

Our calculations also reveal that $V_\mathrm{O}^{+}$ possesses a spin-1/2 ground state and that the negatively charged $(V_\mathrm{Ca}-V_\mathrm{O})^{-}$ complex is likewise paramagnetic, suggesting that these defects may act as sources of magnetic noise in CaO-based quantum defects. Finally, by combining SCAN-relaxed geometries with one-shot HSE calculations within the Franck–Condon framework, we find that several experimentally observed optical absorption and emission bands can be assigned to vacancy complexes as well as the isolated oxygen vacancies. 

These results provide a consistent microscopic understanding of native defects in CaO and establish a foundation for future studies of defect engineering and quantum-defect applications.

\begin{acknowledgments}

This work was supported by the government of the Republic of Korea (MSIT) and the National Research Foundation of Korea (RS-2024-00442710, RS-2025-16068832, RS-2025-25443942) and the National Supercomputing Center with supercomputing resources including technical support (Grant No. KSC-2024-CRE-0395, KSC-2025-CRE-0596).


\end{acknowledgments}



\bibliography{CAO}

@article{wu2020first,
  title={First-principles calculations of oxygen vacancy in CaO crystal},
  author={Wu, Kaili and Liu, Tingyu and Sun, Ruxi and Song, Jiamei and Shi, Chunyu},
  journal={The European Physical Journal D},
  volume={74},
  number={10},
  pages={209},
  year={2020},
  publisher={Springer}
}

@inproceedings{albuquerque2008structural,
  title={Structural, electronics and optical properties of CaO},
  author={Albuquerque, EL and Vasconcelos, MS},
  booktitle={Journal of Physics: Conference Series},
  volume={100},
  number={4},
  pages={042006},
  year={2008}
}

@article{franckel2025absorption,
  title={Absorption lines of F centers in MgO and CaO through time-dependent hybrid-functional theory calculations},
  author={Franckel, Mathilde LD and Falletta, Stefano and Chen, Wei and Pasquarello, Alfredo},
  journal={Physical Review B},
  volume={112},
  number={10},
  pages={104101},
  year={2025},
  publisher={APS}
}

@article{elfimov2002possible,
  title={Possible path to a new class of ferromagnetic and half-metallic ferromagnetic materials},
  author={Elfimov, IS and Yunoki, S and Sawatzky, GA},
  journal={Physical review letters},
  volume={89},
  number={21},
  pages={216403},
  year={2002},
  publisher={APS}
}

@article{osorio2006magnetism,
  title={Magnetism without magnetic ions: percolation, exchange, and formation energies<? format?> of magnetism-promoting intrinsic defects in CaO},
  author={Osorio-Guill{\'e}n, Jorge and Lany, Stephan and Barabash, SV and Zunger, Alex},
  journal={Physical review letters},
  volume={96},
  number={10},
  pages={107203},
  year={2006},
  publisher={APS}
}

@article{kanai2022generalized,
  title={Generalized scaling of spin qubit coherence in over 12,000 host materials},
  author={Kanai, Shun and Heremans, F Joseph and Seo, Hosung and Wolfowicz, Gary and Anderson, Christopher P and Sullivan, Sean E and Onizhuk, Mykyta and Galli, Giulia and Awschalom, David D and Ohno, Hideo},
  journal={Proceedings of the National Academy of Sciences},
  volume={119},
  number={15},
  pages={e2121808119},
  year={2022},
  publisher={National Academy of Sciences}
}

@article{davidsson2024discovery,
  title={Discovery of atomic clock-like spin defects in simple oxides from first principles},
  author={Davidsson, Joel and Onizhuk, Mykyta and Vorwerk, Christian and Galli, Giulia},
  journal={Nature Communications},
  volume={15},
  number={1},
  pages={4812},
  year={2024},
  publisher={Nature Publishing Group UK London}
}

@article{yuan2024first,
  title={First-principles study of defects and doping limits in CaO},
  author={Yuan, Zhenkun and Hautier, Geoffroy},
  journal={Applied Physics Letters},
  volume={124},
  number={23},
  year={2024},
  publisher={AIP Publishing}
}

@article{chen1990effect,
  title={Effect of substitutional hydride ions on the charge states of oxygen vacancies in thermochemically reduced CaO and MgO},
  author={Chen, Y and Orera, VM and Gonzalez, R and Williams, RT and Williams, GP and Rosenblatt, GH and Pogatshnik, GJ},
  journal={Physical Review B},
  volume={42},
  number={2},
  pages={1410},
  year={1990},
  publisher={APS}
}

@article{blochl1994projector,
  title={Projector augmented-wave method},
  author={Bl{\"o}chl, Peter E},
  journal={Physical review B},
  volume={50},
  number={24},
  pages={17953},
  year={1994},
  publisher={APS}
}

@article{kresse1993ab,
  title={Ab initio molecular dynamics for open-shell transition metals},
  author={Kresse, Georg and Hafner, JJPRB},
  journal={Physical Review B},
  volume={48},
  number={17},
  pages={13115},
  year={1993},
  publisher={APS}
}

@article{sun2015strongly,
  title={Strongly constrained and appropriately normed semilocal density functional},
  author={Sun, Jianwei and Ruzsinszky, Adrienn and Perdew, John P},
  journal={Physical review letters},
  volume={115},
  number={3},
  pages={036402},
  year={2015},
  publisher={APS}
}

@article{freysoldt2014first,
  title={First-principles calculations for point defects in solids},
  author={Freysoldt, Christoph and Grabowski, Blazej and Hickel, Tilmann and Neugebauer, J{\"o}rg and Kresse, Georg and Janotti, Anderson and Van de Walle, Chris G},
  journal={Reviews of modern physics},
  volume={86},
  number={1},
  pages={253--305},
  year={2014},
  publisher={APS}
}

@article{kumagai2023finite,
  title={Finite-size corrections to defect energetics along one-dimensional configuration coordinate},
  author={Kumagai, Yu},
  journal={Physical Review B},
  volume={107},
  number={22},
  pages={L220101},
  year={2023},
  publisher={APS}
}

@article{expform,
  title = {Nist-janaf thermochemical tables 4th ed.},
  author = {M. W. Chase},
  journal = {J. of Physical and Chemical Refernce Data },
  pages = {1529},
  year = {1998},
  month = {Jun},
  }

@article{whited1973exciton,
  title={Exciton thermoreflectance of MgO and CaO},
  author={Whited, RC and Flaten, Christopher J and Walker, WC},
  journal={Solid State Communications},
  volume={13},
  number={11},
  pages={1903--1905},
  year={1973},
  publisher={Elsevier}
}

@article{CaO4.78,
  title={Growth of CaO single crystals from Molten CaCl2 in wet atmosphere},
  author={Ohsato, Hitoshi and Sugimura, Takashi and Kageyama, K},
  journal={Journal of Crystal Growth},
  volume={48},
  number={3},
  pages={459--463},
  year={1980},
  publisher={Elsevier}
}

@article{CaO4.81_1,
  title={Low-temperature thermal expansion of LiH, MgO and CaO},
  author={Smith, Deane K and Leider, HR},
  journal={Applied Crystallography},
  volume={1},
  number={4},
  pages={246--249},
  year={1968},
  publisher={International Union of Crystallography}
}

@article{CaO4.81_2,
  title={The Ca- O (Calcium-Oxygen) system},
  author={Wriedt, HA},
  journal={Bulletin of Alloy phase diagrams},
  volume={6},
  number={4},
  pages={337--342},
  year={1985},
  publisher={Springer}
}

@article{momma2011vesta,
  title={VESTA 3 for three-dimensional visualization of crystal, volumetric and morphology data},
  author={Momma, Koichi and Izumi, Fujio},
  journal={Applied Crystallography},
  volume={44},
  number={6},
  pages={1272--1276},
  year={2011},
  publisher={International Union of Crystallography}
}

@article{henkelman2000climbing,
  title={A climbing image nudged elastic band method for finding saddle points and minimum energy paths},
  author={Henkelman, Graeme and Uberuaga, Blas P and J{\'o}nsson, Hannes},
  journal={The Journal of chemical physics},
  volume={113},
  number={22},
  pages={9901--9904},
  year={2000},
  publisher={American Institute of Physics}
}

@article{vineyard1957frequency,
  title={Frequency factors and isotope effects in solid state rate processes},
  author={Vineyard, George H},
  journal={Journal of Physics and Chemistry of Solids},
  volume={3},
  number={1-2},
  pages={121--127},
  year={1957},
  publisher={Elsevier}
}

@article{heyd2003hybrid,
  title={Hybrid functionals based on a screened Coulomb potential},
  author={Heyd, Jochen and Scuseria, Gustavo E and Ernzerhof, Matthias},
  journal={The Journal of chemical physics},
  volume={118},
  number={18},
  pages={8207--8215},
  year={2003},
  publisher={American Institute of Physics}
}

@article{krukau2006influence,
  title={Influence of the exchange screening parameter on the performance of screened hybrid functionals},
  author={Krukau, Aliaksandr V and Vydrov, Oleg A and Izmaylov, Artur F and Scuseria, Gustavo E},
  journal={The Journal of chemical physics},
  volume={125},
  number={22},
  year={2006},
  publisher={AIP Publishing}
}

@article{cui2015evolution,
  title={Evolution of the electronic structure of CaO thin films following Mo interdiffusion at high temperature},
  author={Cui, Yi and Pan, Yi and Pascua, Leandro and Qiu, Hengshan and Stiehler, Christian and Kuhlenbeck, Helmut and Nilius, Niklas and Freund, Hans-Joachim},
  journal={Physical Review B},
  volume={91},
  number={3},
  pages={035418},
  year={2015},
  publisher={APS}
}

@article{alkauskas2008defect,
  author  = {Alkauskas, Audrius and Broqvist, Peter and Pasquarello, Alfredo},
  title   = {Defect Energy Levels in Density Functional Calculations: Alignment and Band Gap Problem},
  journal = {Phys. Rev. Lett.},
  volume  = {101},
  pages   = {046405},
  year    = {2008}
}

@article{alkauskas2011bandedge,
  author  = {Alkauskas, Audrius and Pasquarello, Alfredo},
  title   = {Band-edge problem in the theoretical determination of defect energy levels: The O vacancy in ZnO as a benchmark case},
  journal = {Phys. Rev. B},
  volume  = {84},
  pages   = {125206},
  year    = {2011}
}

@article{Henderson_1972,
  title = {Temperature Dependence of Luminescence of ${F}^{+}$ and $F$ Centers in CaO},
  author = {Henderson, B. and Chen, Y. and Sibley, W. A.},
  journal = {Phys. Rev. B},
  volume = {6},
  issue = {10},
  pages = {4060--4068},
  numpages = {0},
  year = {1972},
  month = {Nov},
  publisher = {American Physical Society},
  doi = {10.1103/PhysRevB.6.4060},
  url = {https://link.aps.org/doi/10.1103/PhysRevB.6.4060}
}

@article{Welch_1980,
doi = {10.1088/0022-3719/13/9/024},
url = {https://doi.org/10.1088/0022-3719/13/9/024},
year = {1980},
month = {mar},
publisher = {},
volume = {13},
number = {9},
pages = {1791},
author = {L S Welch and A E Hughes and G P Summers},
title = {Photoionisation of the F centre in calcium oxide},
journal = {Journal of Physics C: Solid State Physics}
}

@misc{MC_CWO_2026,
      title={Native defects and erbium impurities in CaWO4}, 
      author={Minseok Choi and Mark E. Turiansky and BaiQing Zhao and Jeff D. Thompson and Chris G. Van de Walle},
      year={2026},
      eprint={2605.24809},
      archivePrefix={arXiv},
      primaryClass={cond-mat.mtrl-sci},
      url={https://arxiv.org/abs/2605.24809}, 
}

@article{AJ_ZnO_2007,
  title = {Native point defects in ZnO},
  author = {Janotti, Anderson and Van de Walle, Chris G.},
  journal = {Phys. Rev. B},
  volume = {76},
  issue = {16},
  pages = {165202},
  numpages = {22},
  year = {2007},
  month = {Oct},
  publisher = {American Physical Society},
  doi = {10.1103/PhysRevB.76.165202},
  url = {https://link.aps.org/doi/10.1103/PhysRevB.76.165202}
}

@article{Medvedeva2007,
  title = {Electronic band structure and carrier effective mass in calcium aluminates},
  author = {Medvedeva, Julia E. and Teasley, Emily N. and Hoffman, Michael D.},
  journal = {Phys. Rev. B},
  volume = {76},
  issue = {15},
  pages = {155107},
  numpages = {6},
  year = {2007},
  month = {Oct},
  publisher = {American Physical Society},
  doi = {10.1103/PhysRevB.76.155107},
  url = {https://link.aps.org/doi/10.1103/PhysRevB.76.155107}
}

@article{subramanian1989dielectric,
  title={Dielectric constants of BeO, MgO, and CaO using the two-terminal method},
  author={Subramanian, MA and Shannon, RD and Chai, BHT and Abraham, MM and Wintersgill, MC},
  journal={Physics and chemistry of minerals},
  volume={16},
  number={8},
  pages={741--746},
  year={1989},
  publisher={Springer}
}

\end{document}